# Quantum Entanglement and the Maximum Entropy States from the Jaynes Principle

A. K. Rajagopal

Naval Research Laboratory, Washington, D. C., 20375-5320.

(Revised version: 25 March 1999)

We show that the Jaynes principle is indeed a proper inference scheme when applied to compound systems and will correctly produce the entangled maximum entropy states compatible with appropriate data. This is accomplished by including the dispersion of the entanglement along with its mean value as constraints and an application of the uncertainty principle. We also construct a "thermodynamic-like" description of the entanglement arising out of the maximum entropy principle.

The importance of understanding quantum entangled states in the field of quantum information theory (quantum teleportation, quantum computation, etc.) has been increasingly recognized. In the quest for quantifying entanglement, the methods of statistical inference of incomplete data have been invoked most recently by Horodecki et. al.[1]. They claimed that one of the frequently used powerful methods is the Jaynes principle which, when applied to composite quantum systems can produce fake entanglement. They suggested a cure based on minimization of entanglement. The purpose of this paper is to show that by incorporating the variance as a second constraint in the Jaynes principle, this problem is resolved. This is physically motivated by the fact that the accuracy of any experimental determination of the mean value is assessed by having its dispersion as small as possible. The clue to our resolution of the problem faced in [1] comes from considerations of the relative Kullback-Leibler entropy and the uncertainty principle deduced from an application of Schwartz inequality to the entanglement operator and the operator arising from its square. The minimization of entanglement in [1] is here found to be nothing other the statement of minimum uncertainty. We have thus combined the two inference schemes of [1], namely the entropic one and the entanglement one, into a single scheme. This idea of using "proper" set of data to obtain physically meaningful inference is entirely within the philosophy espoused by Jaynes himself [2]. We may add that the notion of the use of linearly independent "sufficient statistics" in the theory of statistical inference first propounded by Fisher [3] is closely parallel to the use of mean values of linearly independent observables in constructing completely a state of the quantum system. In view of the importance of the compound quantum systems in many new phenomena involving entangled quantum states in the recent past, we hope that the insight provided in this paper may prove useful in a proper quantification of the concept of entanglement.

We follow [1] and begin by considering the Bell-CHSH observable

$$\hat{B} = 2\sqrt{2}\left(\left|\Phi^+\right\rangle\left\langle\Phi^+\right| - \left|\Psi^-\right\rangle\left\langle\Psi^-\right|\right) \tag{1}$$

with the mean value

$$\left\langle\hat{B}\right\rangle \equiv Tr\hat{\rho}\hat{B} = b, \quad 0 \leq b \leq 2\sqrt{2}, \tag{2}$$

where $\hat{\rho}$ is the system density matrix. We use the Bell basis as in [1]

$$\left|\Phi^m\right\rangle = \frac{1}{\sqrt{2}}\left(\left|\uparrow\uparrow\right\rangle m\left|\downarrow\downarrow\right\rangle\right), \left|\Psi^\pm\right\rangle = \frac{1}{\sqrt{2}}\left(\left|\uparrow\downarrow\right\rangle \pm \left|\downarrow\uparrow\right\rangle\right). \tag{3}$$

Before we apply the Jaynes inference scheme, we make two observations.

First, by using the Kullback-Leibler relative entropy [4] which gives a measure of the difference between two density operators, $\hat{\rho}_1, \hat{\rho}_2$,

$$K(\hat{\rho}_1, \hat{\rho}_2) \equiv Tr\hat{\rho}_2(\ln\hat{\rho}_2 - \ln\hat{\rho}_1) \geq 0, \tag{4}$$

we show that the entropy, $S_2$, associated with the density operator $\hat{\rho}_2$, determined by the Jaynes principle of maximum entropy with two constraints, $b = Tr\hat{B}\hat{\rho}_2$ and another constraint as yet unspecified, is smaller than the entropy $S_1$ associated with the density operator $\hat{\rho}_1$ with only one constraint chosen such that $b = Tr\hat{B}\hat{\rho}_1$: $S_2 \leq S_1$. This is because the Jaynes principle in the one-constraint case gives

$$\hat{\rho}_1 = \left(Z_1(\lambda_1)\right)^{-1}\exp-\left(\lambda_1\hat{B}\right), \quad Z_1(\lambda_1) = Tr\exp-\left(\lambda_1\hat{B}\right),$$
$$S_1 = \lambda_1 b + \ln Z_1(\lambda_1). \tag{5}$$

This gives us a motivation to look for a second constraint in the entanglement problem. It may not be out of place here to mention that there have been entropic measures suggested earlier such as entanglement of formation [5], quantum relative entropy [6], and quantum mutual entropy [7]. These methods have been used to study entanglement features in the Jaynes-Cummings model of interacting two-level system and radiation [8, 9].

Second, any measurement of b must be accompanied by a statement about its dispersion quantifying the accuracy of the result of the measurement. So we consider the square of the Bell-CHSH observable given by Eq.(1):

$$\hat{B}^2 = 8\left(\left|\Phi^+\right\rangle\left\langle\Phi^+\right| + \left|\Psi^-\right\rangle\left\langle\Psi^-\right|\right). \tag{6}$$

This operator is linearly independent of the first one and its expectation value gives the dispersion about the mean value, b:

$$\sigma^2 \equiv Tr\hat{\rho}\,\hat{B}^2, \tag{7}$$

which clearly obeys the obvious inequality $\sigma^2 - b^2 \geq 0$. Since the two operators considered here commute, an application of Schwartz inequality in the form

$$\left\langle\hat{X}^2\right\rangle\left\langle\hat{Y}^2\right\rangle \geq \left\langle\hat{X}\hat{Y}\right\rangle^2 \tag{8}$$

with $\hat{X} = \hat{B}, \hat{Y} = \hat{B}^2$, and the observation that $\hat{X}^2 = \hat{B}^2, \hat{Y}^2 = 8\hat{B}^2, \hat{X}\hat{Y} = 8\hat{B}$, and using the definitions above, gives us an important uncertainty principle, namely

$$\sigma^2 \geq 2\sqrt{2}\, b. \tag{9}$$

The equality in Eq.(9) gives the minimum uncertainty. This development suggests that we employ eq.(6) as the second constraint in this problem.

Now, applying the maximum entropy principle with these two constraints given above, we obtain, after some algebra, the density matrix

$$\hat{\rho}_{J2} = \frac{1}{16}\left(\sigma^2 + 2\sqrt{2}\,b\right)|\Phi^+\rangle\langle\Phi^+| + \frac{1}{16}\left(\sigma^2 - 2\sqrt{2}\,b\right)|\Psi^-\rangle\langle\Psi^-| + \\ + \frac{1}{2}\left(1 - \frac{\sigma^2}{8}\right)\left(|\Phi^-\rangle\langle\Phi^-| + |\Psi^+\rangle\langle\Psi^+|\right). \tag{10}$$

Applying the separability criterion given in [1] that the eigenvalues of this density matrix do not exceed 1/2, we have the inequalities

$$\frac{1}{16}\left(\sigma^2 \pm 2\sqrt{2}\,b\right), \quad \frac{1}{2}\left(1 - \frac{\sigma^2}{8}\right) \leq \frac{1}{2}. \tag{11}$$

Thus if

$$\sigma^2 \rangle \left(8 - 2\sqrt{2}\,b\right), \tag{12}$$

the state $\hat{\rho}_{J2}$ is inseparable.

We now make several observations:

(1) The minimum entanglement state obeying the Jaynes maximum entropy principle obtained in [1] corresponds here to the minimum uncertainty state given by $\sigma^2_{min} = 2\sqrt{2}\,b$. In this way, the two inference schemes in [1] gets combined into one following the Jaynes principle. The inseparability condition in Eq.(12) becomes $b \rangle \sqrt{2}$.

(2) For the single constraint case considered in [1], the dispersion is found to be

$$\sigma^2_{J1} = 4\left(1 + \frac{b^2}{8}\right), \tag{13}$$

The obvious inequality $\sigma^2 - b^2 \geq 0$ then leads to the condition that b should obey the inequality $0 \langle b \leq 2\sqrt{2}$.

(3) The dispersion in the one-constraint state given in [1] is found to be larger than the minimum dispersion, as expected:

$$\sigma^2_{J1} - \sigma^2_{min} = 4\left(1 + \frac{b^2}{8}\right) - 2\sqrt{2}\,b = \frac{1}{2}\left(b - 2\sqrt{2}\right)^2 \geq 0. \tag{14}$$

(4) One obtains a pure state, $\hat{\rho} = |\Phi^+\rangle\langle\Phi^+|$, if we are at the minimum allowed values, $\sigma^2 = 8, \quad b = 2\sqrt{2}$. Thus the purification of the state is achieved by a suitable choice of the values for the constraints, which arise from general considerations of uncertainty principle and the Bell equality.

(5) A "thermodynamic-like" version of the above expression for the density matrix and the corresponding von Neumann entropy may be developed as follows. We first express the density matrix in the form

$$\hat{\rho} = Z^{-1}(\lambda_1, \lambda_2) \begin{Bmatrix} \exp(-\lambda_1 2\sqrt{2} - 8\lambda_2) |\Phi^+\rangle\langle\Phi^+| \\ +\exp(\lambda_1 2\sqrt{2} - 8\lambda_2) |\Psi^-\rangle\langle\Psi^-| \\ +(|\Phi^-\rangle\langle\Phi^-| + |\Psi^+\rangle\langle\Psi^+|) \end{Bmatrix}, \quad (15)$$

$$Z(\lambda_1, \lambda_2) = \left\{ \exp(-\lambda_1 2\sqrt{2} - 8\lambda_2) + \exp(\lambda_1 2\sqrt{2} - 8\lambda_2) + 2 \right\}.$$

The Lagrange parameters are determined by the usual relations,

$$b = -\frac{\partial \ln Z(\lambda_1, \lambda_2)}{\partial \lambda_1}, \quad and \quad \sigma^2 = -\frac{\partial \ln Z(\lambda_1, \lambda_2)}{\partial \lambda_2}. \quad (16)$$

The von Neumann entropy is then found to be

$$S = \ln Z(\lambda_1, \lambda_2) + \lambda_1 b + \lambda_2 \sigma^2. \quad (17)$$

We may interpret $\ln Z(\lambda_1, \lambda_2)$ as the "Free energy" of the Bell-CHSH state. When both the Lagrange multipliers go to minus infinity, the entropy vanishes for $\sigma^2 = 8$, $b = 2\sqrt{2}$, and the Bell-CHSH pure state is reached, at this "zero temperature" limit, if we identify the Lagrange parameters as in the usual statistical mechanics, $\lambda_1 = -\beta$, $\lambda_2 = -\beta\mu$, where $\beta$ is identified with the inverse "temperature" and $\mu$ with the "chemical potential". In fact solving for the Lagrange multipliers in terms of the constraint patrameters, at equilibrium, we have,

$$\lambda_1 = -\frac{1}{4\sqrt{2}} \left\{ \ln(\sigma^2 + b2\sqrt{2}) - \ln(\sigma^2 - b2\sqrt{2}) \right\},$$

$$\lambda_2 = -\frac{1}{16} \left\{ \ln(\sigma^2 + b2\sqrt{2}) + \ln(\sigma^2 - b2\sqrt{2}) - 2\ln(8 - \sigma^2) \right\}. \quad (18)$$

and these certainly seem to admit of the "thermodynamic" identification suggested above. Since all the quantities appearing here pertain to entanglement, we may say that this development lends itself to the notion of "thermodynamics of entanglement".

In conclusion, we have shown here that if the mean value is going to be the measaured quantity, the theory should be so constructed such that the fluctuation about this mean is minimal. In this regard, it seems reasonable that the Jaynes principle applied to the problem of quantum entanglement should involve both the mean value and the dispersion of the Bell operator as constraints. We hope that the present work has thus clarified the Jaynes scheme of statistical inference for entanglement processing. This clarification is particularly important in view of the power of the Jaynes principle in solving important statistical inference problems, be it classical or quantal [2]. We also outline an interpretation of the results obtained here in terms of the thermodynamic language by identifying the Lagrange multipliers in terms of inverse "temperature" and "chemical potential". Thus it appears that the maximum entropy principle may lead to the notions of "thermodynamics of entanglement" that is being discussed in the current literature [10].

Thanks are due to Professors R. Horodecki and Sumuyoshi Abe for reading an early draft of this paper. Professor R. Horodecki also provided me with a copy of the last reference to their work on thermodynamical analogies in [10].This work is supported in part by the Office of Naval Research.